\documentclass[prd,twocolumn,nofootinbib]{revtex4}
\usepackage{bm}
\usepackage{amssymb}
\usepackage{graphicx}
\usepackage{epstopdf}
\usepackage{mathrsfs}
\usepackage{amsmath}

\usepackage[utf8]{inputenc}
\usepackage[english]{babel}
\usepackage{xcolor}

\begin{document}
\title{Scalar and gravitational hair for extreme Kerr black holes} 
\author{Lior M.~Burko$^1$, Gaurav Khanna$^2$, and Subir Sabharwal$^{2}$
}
\affiliation{
$^1$ Theiss Research, La Jolla, California 92037, USA \\ 
$^2$ Department of Physics, University of Massachusetts, Dartmouth, Massachusetts  02747, USA
}
\date{May 13, 2020}
\begin{abstract} 

For scalar perturbations of an extreme Reissner-Nordstr{\" o}m black hole we show numerically that the Ori pre-factor equals the Aretakis conserved charge. We demonstrate a linear relation of a generalized Ori pre-factor -- a certain expression obtained from the late-time expansion or the perturbation field at finite distances -- and the Aretakis conserved charge for a family of scalar or gravitational perturbations of an extreme Kerr black hole, whose members vary only in the radial location of the center of the initial packet. We infer that it can be established that there is an Aretakis conserved charge for scalar or gravitational perturbations of extreme Kerr black holes. This conclusion, in addition to the calculation of the Aretakis charge, can be made from measurements at a finite distance: Extreme Kerr black holes have gravitational hair that can be measured at finite distances. This gravitational hair can in principle be detected by 
gravitational-wave detectors. 

\end{abstract}
\maketitle


{\it Introduction and summary.} 
Extreme spherically symmetric and charged black holes [extreme Reissner-Nordstr{\" o}m black holes (BHs), hereafter ERN] have been shown to carry massless 
scalar hair that can be measured at future null infinity (${\mathscr{I}^+}$) \cite{aretakis}. This scalar hair is a certain quantity $s[\psi]$ which is evaluated at ${\mathscr{I}^+}$ and which equals the Aretakis charge, a non-vanishing quantity $H[\psi]$ which is calculated on the BH's event horizon (EH, ${\mathscr{H}^+}$) but vanishes if the BH is non-extreme. 

Since the scalar hair at ${\mathscr{I}^+}$ is intimately related to the Aretakis conserved charge on ${\mathscr{H}^+}$, one may suspect that corresponding conserved charges for other fields on either ERN or extreme Kerr (EK) BHs may also be related to observable hair at ${\mathscr{I}^+}$, or be measurable at finite distances. Specifically, conserved Aretakis charges were found in ERN, in addition for massless scalar fields \cite{aretakis-2012} also for massive scalar fields, for coupled linearized gravitational and electromagnetic fields \cite{lucietti-2012}, for charged scalar perturbations \cite{zimmerman-2017}, and in EK for scalar \cite{aretakis-2012}, electromagnetic, and gravitational perturbations \cite{lucietti-reall-2012,burko-khanna-2017,gralla-2018}. 

Ori showed  that the Aretakis charge can also be used in order to determine a certain pre-factor $e[\psi]$ in the late time expansion of scalar field perturbation fields in ERN as measured at a finite distance  \cite{ori-2013}. Here, we first show numerically that for scalar perturbations of ERN the Ori pre-factor $e[\psi]$ equals $H[\psi]$, and therefore can be used in order to measure the Aretakis conserved charge at a finite distance. It follows that $e[\psi]$ can be interpreted as scalar hair measured outside the BH. 

We then go beyond the framework of scalar perturbations of ERN to EK, and show numerically that analogous pre-factors can be formulated also for scalar and gravitational perturbations of EK. Since the value of the Aretakis charge depends on the initial data of the perturbation field, it follows that information on the preparation of the perturbation field can be inferred at great distances from the BH measurements, in apparent contradiction of the established no-hair theorem \cite{bekenstein1,bekenstein2,bekenstein3}. That is, we bring evidence that 
in addition to the three externally observable classical parameters, specifically the BH's mass $M$, charge $q$, and spin angular momentum $a$, it is in principle possible to also detect with a gravitational-wave detector the gravitational Aretakis charge of EK.  

{\it Setting up the problem.} 
Following Ori \cite{ori-2013} we write the late time expansion of a field $\psi_{s,\ell,m}^{\rm I}$ as 
\begin{eqnarray}
\psi_{s,\ell,m}^{\rm I}(t,r,\theta)&=&e_{s,\ell,m}^{\rm I}\,r(r-M)^{-p^{\rm I}_{s,\ell,m}}t^{-n^{\rm I}_{s,\ell,m}}\,\Theta_{s,\ell,m}^{\rm I}(\theta)
\nonumber \\
&+&O(t^{-n^{\rm I}_{s,\ell,m}-k_{s,\ell,m}^{\rm I}})
\label{LTE}
\end{eqnarray}
in Boyer-Lindquist coordinates, where $s$ is the field's spin, $\ell,m$ are the spherical harmonic numbers, and the index ${\rm I}$ corresponds to the BH type, i.e., ${\rm I}=\{{\rm ERN,EK}\}$. Here, $e_{s,\ell,m}^{\rm I}$ is a generalized Ori pre-factor. The case studied in \cite{ori-2013} corresponds to $\psi_{0,\ell,0}^{\rm ERN}$, for which it was found that $e_{0,\ell,0}^{\rm I}=(-4)^{\ell+1}M^{3\ell+2}e$ \cite{ori-2013}, where $e[\psi]$ is a certain pre-factor that depends on the initial data (and which is given explicitly in \cite{ori-2013}), and $p^{\rm ERN}_{0,\ell,0}=\ell+1$, $n^{\rm ERN}_{0,\ell,0}=2\ell+2$, and $\Theta^{\rm ERN}_{0,0,0}(\theta)=1$. The late-time expansion (\ref{LTE}) is expected to be valid for $t\gg r_*$, where $r_*$ is the tortoise coordinate. Specifically, we may expect $r$-dependent correction terms when this condition is not satisfied. Comparing \cite{aretakis} and \cite{ori-2013} we expect that $e^{\rm ERN}_{0,0,0}[\psi]=-4M^2H[\psi]$.

{\it Numerical approach.} To test this prediction, and to set up the framework for generalization to EK and to gravitational perturbations, we write the 2+1 Teukolsky equation in ERN or EK backgrounds for azimuthal ($m=0$) modes in compactified hyperboloidal coordinates ($\tau,\rho,\theta,\varphi$), such that $\mathscr{I}^+$ is included in the computational domain at a finite radial (in $\rho$) coordinate~\cite{Zenginoglu:2007jw}. We re-write the second-order hyperbolic partial differential equation  as a coupled system of two first-order hyperbolic equations. We solve this system for the scalar field case by implementing a second-order Richtmeyer-Lax-Wendroff iterative evolution scheme~\cite{Zenginoglu:2011zz, Burko:2016uvr}. For the gravitational case we implement a sixth-order (in $\rho$) WENO (Weighted Essentially Non-Oscillatory) finite-difference scheme with explicit time-stepping \cite{burko-khanna-2017}. These codes converge with second-order temporally and angularly.

The initial data are a compactly supported ``truncated'' gaussian with non-zero initial field values on ${\mathscr{H}^+}$, but similar results  are expected also for other forms of initial data. Specifically, in hyperboloidal coordinates $(\rho,\tau)$ (see \cite{Burko:2016uvr} for definitions), the initially spherical ($\ell=0$) Gaussian pulse is
non-vanishing in the range $\rho/M\in\left[0.95, 8\right]$, has a width of $0.1M$ and centered close to the BH (at $\rho/M=1.0$, $1.1$, $1.2$, $1.3$, $1.4$ and $1.5$ respectively). (The EH, ${\mathscr{H}^+}$, is at $\rho=0.95M$ for ERN and EK in these coordinates.) The outer boundary is located at $S=\rho(\mathscr{I}^+)=19.0M$. 

The computations were performed on IBM 32-core Power9 servers accelerated by Nvidia V100 GPGPUs. Our resolution for each production run  was $\,\Delta\rho=M/6,400$, $\,\Delta\tau=M/12,800$, $\,\Delta\theta=\pi/64$, which we run in quadrupole precision (128-bit, i.e., to $\sim30$ decimal digits). The combination of quadruple-precision floating point numerics and the extremely high-resolution resulted in computationally intensive simulations, which took two weeks for each run to get to $t/M\sim 1,600$. 

\begin{figure}[h]
\includegraphics[width=8.5cm]{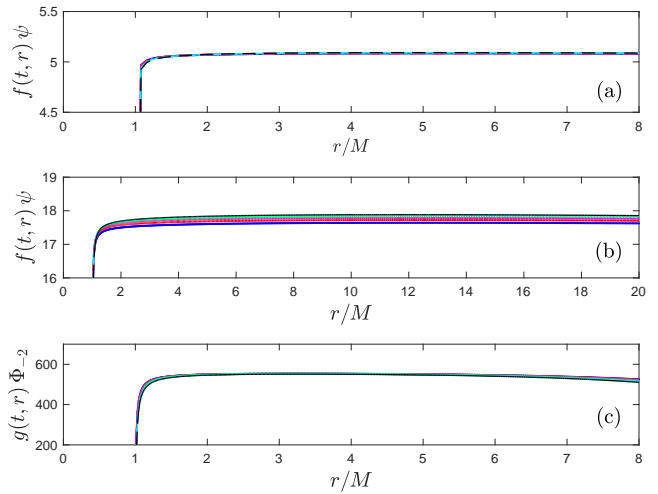}
\caption{The values of $e^{\rm I}_{s,\ell,0}[\psi](t)$ as functions of $r/M$. These values are shown for the data set for which at the gaussian's center $\rho/M=1.0$. Top panel (a): ERN with $s=0,\ell=0$. Middle panel (b): EK with $s=0,\ell=0$. Bottom panel (c): EK with $s=-2,\ell=2$. The values are plotted for $t/M=1100$, $1200$, $1300$, $1400$, $1500$, and $1600$. [For panel (c) the time value was replaced with $1553$.] The function $f(t,r)=(t/M)^2(1-M/r)$ and the function $g(t,r)=M(t/M)^6(r/M)^4(1-M/r)^5$. }
\label{psi_r}
\end{figure}

{\it Scalar perturbations of ERN.} 
We calculate $e^{\rm ERN}_{0,0,0}[\psi]$  directly from Eq.~(\ref{LTE}), and $H^{\rm ERN}_{0,0,0}[\psi]$ from 
\begin{equation}\label{H_spherical}
H^{\rm I}_{0,0,0}[\psi]=-\frac{M^2}{4\pi}\int_{\mathscr{H}^+}\,\partial_r(r\psi)\,d\Omega\, ,
\end{equation}
where ${\rm I}={\rm ERN}$. 
To determine $e^{\rm ERN}_{0,0,0}[\psi]$ we calculate it for a set of finite values of the time. Figure \ref{psi_r}(a) shows $e^{\rm ERN}_{0,0,0}[\psi]$ at a number of time values as a function of the Schwarzschild coordinate $r$, for the initial data set for which the gaussian is centered at $\rho/M=1.0$. Notice that the numerical constancy of $(t/M)^2(1-M/r)\psi_{0,0,0}^{\rm ERN}$ for small values of $r/M$ suggests that $p_{0,0,0}^{\rm ERN}=1$ and $n_{0,0,0}^{\rm ERN}=2$, as expected from \cite{ori-2013}. For larger values of $r/M$ the constant value starts to vary, as expected from the expansion of \cite{ori-2013}. Equation (\ref{LTE}) suggests that $e^{\rm ERN}_{0,0,0}[\psi](t)$ is time dependent, and that when 
$(t/M)^2(1-M/r)\psi_{0,0,0}^{\rm ERN}$ is plotted as a function of inverse time, the value of $k$ can be determined. We see in Fig.~\ref{psi_r} that there is indeed time dependence as expected.

\begin{figure}[h]
\includegraphics[width=8.5cm]{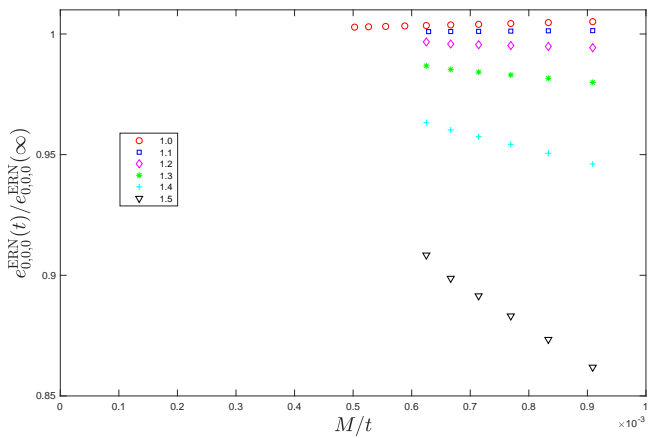}
\caption{The values of $e^{\rm ERN}_{0,0,0}[\psi](t)$, normalized by their values as $t\to\infty$, as functions of $M/t$. These values are shown for each initial data set, parametrized by the $\rho/M$ value at the center of the gaussian packet.}
\label{e_time}
\end{figure}

The time dependence of $e^{\rm ERN}_{,0,0}[\psi](t)$ is shown in greater detail in Fig~\ref{e_time}, which displays for each initial data set the values of $e^{\rm ERN}_{0,0,0}[\psi](t)$. We then extrapolate the values to $M/t\to 0$ by fitting to a linear function and finding the intercept and the slope to determine $e^{\rm ERN}_{0,0,0}[\psi]$. The linearity suggests that $k_{0,0,0}^{\rm ERN}=1$, in agreement with \cite{ori-2013}.

\begin{figure}[h]
\includegraphics[width=8.5cm]{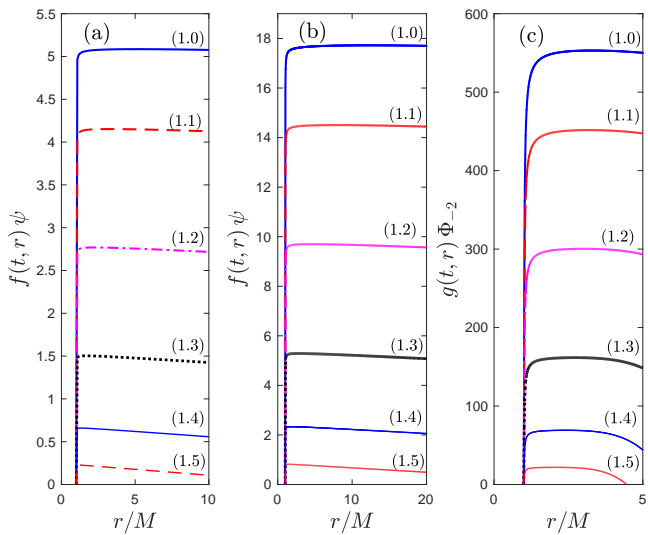}
\caption{The values of $e^{\rm I}_{s,\ell,0}[\psi](t/M=1500)$ as functions of $r/M$, shown for each initial data set, parametrized by the $\rho/M$ value at the center of the gaussian packet. Left panel (a): ERN with $s=0,\ell=0$. Center panel (b): EK with $s=0,\ell=0$. Right panel (c): EK with $s=-2,\ell=2$. }
\label{psi_v_r_1500}
\end{figure}

The values of $e^{\rm ERN}_{0,0,0}[\psi]$ depend on the choice of the initial data set. In Fig.~\ref{psi_v_r_1500}(a) we show $(t/M)^2(1-M/r)\psi_{0,0,0}^{\rm ERN}$ for each initial data set as functions of $r/M$. As the center of the initial gaussian packet moves outward (to larger $\rho$ values) the value of $(t/M)^2(1-M/r)\psi_{0,0,0}^{\rm ERN}$ decreases. 

\begin{figure}[h]
\includegraphics[width=7.5cm]{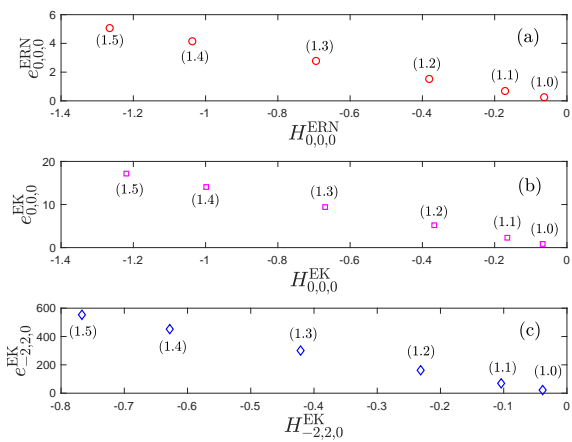}
\caption{The pre-factor $e^{\rm I}_{s,\ell,0}[\psi]$  shown as a function of the Aretakis charge $H^{\rm I}_{s,\ell,0}[\psi]$ for the different initial data sets (parametrized with $\rho/m$ at the center of the gaussian initial packet). Top panel (a): ERN with $s=0,\ell=0$. Middle panel (b): EK with $s=0,\ell=0$. Bottom panel (c): EK with $s=-2,\ell=2$.}
\label{eH}
\end{figure}

Finally, Fig.~\ref{eH}(a) shows the values of $e^{\rm ERN}_{0,0,0}[\psi]$ as a function of the corresponding $H^{\rm ERN}_{0,0,0}[\psi]$ for the different data sets. 
Fitting our numerical data to $e^{\rm ERN}_{0,0,0}[\psi]=\alpha\,H^{\rm ERN}_{0,0,0}[\psi]+\beta$ we find that $\alpha=-4.0024\pm 0.0013$ and $\beta=(1.8\pm 9.6)\times 10^{-4}$, consistently with our expectation. The Ori pre-factor $e$ equals the Aretakis charge $H$. 

{\it Scalar perturbations of EK.} We next extend the analysis from the case of a scalar field in ERN to scalar and gravitational perturbations of EK. First, we set up the initial value problem for scalar field perturbations similarly as for ERN. We use the expansion \ref{LTE} as an Ansatz. The results for the scalar case in EK are shown in Figs.~\ref{psi_r}(b), \ref{psi_v_r_1500}(b), and \ref{eH}(b). These result suggest that Eq.~(\ref{LTE}) describes well also the field for this case. Fitting the parameters to this Ansatz, we find that $p^{\rm EK}_{0,0,0}=1$ and $n^{\rm EK}_{0,0,0}=2$. We also find that $\Theta^{\rm EK}_{0,0,0}(\theta)=1$. To find $H^{\rm EK}_{0,0,0}[\psi]$ we again use Eq.~(\ref{H_spherical}) with ${\rm I}={\rm EK}$. 
Seeking a linear relation of the form $e^{\rm EK}_{0,0,0}[\psi]=\alpha\,H^{\rm EK}_{0,0,0}[\psi]+\beta$ we find that $\alpha=-14.13\pm 0.03$ and $\beta=-0.048\pm 0.023$. The linear relation of $e^{\rm EK}_{0,0,0}[\psi]$ and $H^{\rm EK}_{0,0,0}[\psi]$ suggest that also in this case the Aretakis conserved charge can be measured at a finite distance, and that a generalized Ori pre-factor can be used in order to measure it.

{\it Gravitational perturbations of EK.} Finally, we consider EK gravitational perturbations with $s=-2$ and $\ell=2$. We write the Teukolsky equation for a Kerr BH with parameters $M,a$ for the variable $\Phi_{-2}$, which is related to the Teukolsky function $\Psi_{-2}^{\rm K}$ in the Kinnersley tetrad and Boyer-Lindquist coordinates via 
$\Phi_{-2}=(r/\,\Delta^2)\,\Psi_{-2}^{\rm K}$, where $\Delta=r^2-2Mr+a^2$. Since the Weyl scalar $\psi_4^{\rm HH}$ in the Hartle-Hawking tetrad is related to its Kinnersley tertrad counterpart,  $\psi_4^{\rm K}$, via a type-III transformation, or  $\psi_4^{\rm HH}=4(r^2+a^2)^2\,\Delta^{-2}\, \psi_4^{\rm K}$ \cite{poisson-2004} and that $\Psi_{-2}^{\rm K}=(r-ia\,\cos\theta)^4\,\psi_4^{\rm K}$ \cite{Teukolsky} we find that  
\begin{equation}
\Phi_{-2}=\frac{r\left(r-ia\,\cos\theta\right)^4}{4\left(r^2+a^2\right)^2}\,\psi_4^{\rm HH}\, ,
\end{equation}
and use $\Phi_{-2}$ with $\ell=2,m=0$ and $a=M$ for $\psi^{\rm EK}_{-2,2,0}$. Note that at great distances, as $r\ggg M$, $\psi^{\rm EK}_{-2,2,0}\sim(r/4)\psi_4^{\rm HH}\sim r\psi_4^{\rm K}$. Therefore, determination of  $\psi^{\rm EK}_{-2,2,0}$ at great distances allows us to measure directly the Weyl scalar $\psi_4^{\rm K}$ in the Kinnersley tetrad.  Conversely, measurement with a gravitational wave detector at a great distance of $\psi_4^{\rm K}$ allows us to calculate $\psi^{\rm EK}_{-2,2,0}$ if the distance to the source is known. 

\begin{figure}[]
\includegraphics[width=7.5cm]{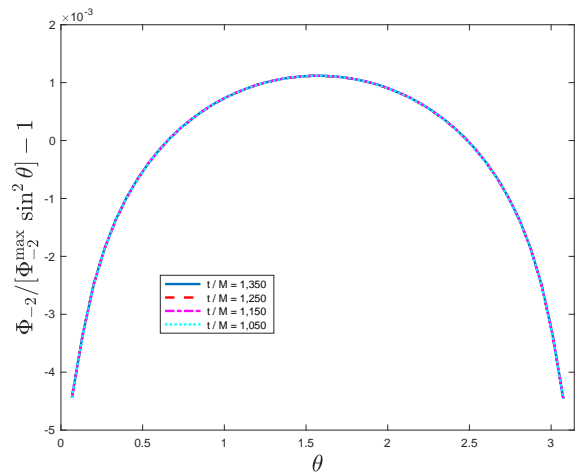}
\caption{The relative difference of the Weyl scalar $\psi_4$ (normalized by its maximal value) and $\Theta(\theta)=\,\sin^2\theta$ as a function on the polar angle $\theta$ at a fixed value of $\rho/M=2$ for four different time values, $t/M=1,050$ (dotted), 1,150 (dash-dotted), 1,250 (dashed), and 1,350 (solid). On the scale shown these plots cannot be resolved.}
\label{angle}
\end{figure}

We plot $\Phi_{-2}$  for a fixed $\rho$ as a function of $\theta$ for a set of $\tau$ values in Fig.~\ref{angle}. Since our angular resolution is $\,\Delta\theta=\pi/64$ and our code converges angularly with second order, we would expect our angular numerical error to be (a few)$\times 10^{-3}$. We find that the angular function $\Theta(\theta)$ deviates from $\,\sin^2\theta$ by no more than (a few)$\times 10^{-3}$. 
Therefore, we could not distinguish numerically between our numerical function $\Theta(\theta)$ and $\,\sin^2\theta$. 

We calculate $e^{\rm EK}_{-2,2,0}[\psi]$  directly from Eq.~(\ref{LTE}), and motivated by  \cite{lucietti-2012}, we calculate $H^{\rm EK}_{-2,2,0}[\psi]$ by 
\begin{equation}
H^{\rm EK}_{-2,2,0}[\psi]=-\frac{8}{3\pi}M^2\int_{\mathscr{H}^+}\,\partial_r\Phi_{-2}\,d\Omega\, .
\end{equation}
(Note that $\psi_4$ decays to 0 at late times on $\mathscr{H}^+$.) We only calculate here the real part of $\psi_4$: Because of the linearity of the Teukolsky equation we can always perform a Wick rotation, and obtain commensurate results for the imaginary part. 

The results for the Weyl scalar $\psi_4$ are shown in Figs.~\ref{psi_r}(c), \ref{psi_v_r_1500}(c), and \ref{eH}(c). Again, we find that the Ansatz (\ref{LTE}) describes the field behavior well. Fitting the parameters to this Ansatz, we find that $p^{\rm EK}_{-2,2,0}=5$ and $n^{\rm EK}_{-2,2,0}=6$. Seeking a linear relation of the form $e^{\rm EK}_{-2,2,0}[\psi]=\alpha\,H^{\rm EK}_{-2,2,0}[\psi]+\beta$ we find that $\alpha=-729.7\pm 0.6$ and $\beta=-6.3\pm 0.3$. The linear relation of $e^{\rm EK}_{-2,2,0}[\psi]$ and $H^{\rm EK}_{-2,2,0}[\psi]$ suggest that also in this case the Aretakis conserved charge can be measured at a finite distance, and that a generalized Ori pre-factor can be used in order to measure it. We summarize our results in Table \ref{table}.

\begin{table}[]
\begin{tabular}{ || l |c|c||c|c|c|c|c || }
\hline
$I$ & $s$ & $\ell$ & $p$ & $n$ & $\Theta(\theta)$ & $\alpha$ & $\beta$ \\
\hline
\hline
ERN &  0 & 0 & 1 & 2 & 1 & $-4.0024\pm 0.0013$ & $(1.8\pm 9.6)\times 10^{-4}$ \\
\hline
EK & 0 & 0 & 1 & 2  & 1 & $-14.13\pm 0.03$ & $-0.048\pm 0.023$ \\
\hline
EK &  -2 & 2 & 5 & 6 & $\,\sin^2\theta$ & $-729.7\pm 0.6$ & $-6.3\pm 0.3$ \\
\hline
\end{tabular}
\caption{The parameters used in the expansion (\ref{LTE}), and the fitted parameters $\alpha,\beta$ in the linear relation $e^{\rm I}_{s,\ell,0}[\psi]=\alpha\,H^{\rm I}_{s,\ell,0}[\psi]+\beta$. }
\label{table}
\end{table}

{\it Discussion.} 
The values for the Ori pre-factor, and therefore also for the Aretakis charge -- when compared between members of the same initial data family which differ from each other just by the distance of the center of the initial packet -- are suggested by our results to be universal, i.e., they depend only weakly on the spin of the field and on whether the BH is ERN or EK (Fig.~\ref{psi_v_r_1500}). 

The linear relation of the Ori pre-factor and the Aretakis conserved charge for either scalar or gravitational perturbations of EK suggests that we could make measurements at a finite distance and conclude that the BH has a conserved charge, and therefore establish also that it is an extreme BH. Moreover, by using the (numerically determined) value of the parameter $\alpha$ (or, in the case of scalar perturbations of ERN, its analytical value) we can calculate the value of the Aretakis charge. If the measured quantity appears to behave as for an ERN or EK for some time, and then decays as for a non-extreme BH (i.e., it is a transient behavior), we can establish that it is a nearly extreme BH (see also \cite{BKS-2019}). Since the value of the Aretakis charge depends on the perturbation field (cf.~Fig.~\ref{psi_v_r_1500}), and this value can be found from observations at a finite distance, this is a procedure for detecting gravitational hair of EK. 

Extreme Kerr BHs that are perturbed gravitationally have hair, and this determination and also the calculation of the strength of the hair can be made at finite distances by measuring the Weyl scalar $\psi_4$ directly from the gravitational wave strain. Specifically, gravitational wave detectors can be used to measure this gravitational-field hair of extreme black holes. 

This apparent contradiction of the no-hair theorem pertains to extreme BHs, which require fine tuning of the astrophysical processes that created them. Realistic BHs are more likely to be nearly extreme, and therefore would present transient hair that could in principle be detected by gravitational-wave detectors. 

Work on higher-$\ell$ modes and non-azimuthal ($m\ne 0$) modes is currently underway. 
Measurement of gravitational hair of EK at $\mathscr{I}^+$ awaits further work. 


{\it Acknowledgements.}  The authors thank Shahar Hadar and Achilleas Porfyriadis for discussions. S.S.~ thanks the University of Massachusetts, Dartmouth for hospitality duration the performance of this work. 
Many of the computations were performed on the MIT/IBM Satori GPU supercomputer supported by the Massachusetts Green High Performance Computing Center (MGHPCC). G.K.~acknowledges research support from NSF Grants No. PHY-1701284 and No. DMS-1912716 and 
Office of Naval Research/Defense University Research Instrumentation Program
(ONR/DURIP) Grant No. N00014181255.

\end{document}